\documentclass[twocolumn,showpacs,preprintnumbers,aps,prl,amsmath,amssymb,floatfix]{revtex4}

\usepackage{graphicx}% Include figure files
\usepackage{dcolumn}% Align table columns on decimal point
\usepackage{bm}% bold mat
\usepackage{mathrsfs}

\begin{document}

\title{Direct absorption imaging of ultracold polar molecules}

\author{D. Wang, B. Neyenhuis, M. H. G. de Miranda, K.-K. Ni,* S. Ospelkaus,** D. S. Jin, and J. Ye}
\affiliation{JILA, National Institute of Standards and Technology and University of Colorado,
Department of Physics, University of Colorado, Boulder, CO 80309-0440, USA}
\date{\today}

\begin{abstract}
We demonstrate a scheme for direct absorption imaging of an ultracold ground-state polar molecular gas near quantum degeneracy. A challenge in imaging molecules is the lack of closed optical cycling transitions. Our technique relies on photon shot-noise limited absorption imaging on a strong bound-bound molecular transition. We present a systematic characterization of this imaging technique. Using this technique combined with time-of-flight (TOF) expansion, we demonstrate the capability to determine momentum and spatial distributions for the molecular gas. We anticipate that this imaging technique will be a powerful tool for studying molecular quantum gases.
\end{abstract}


%\pacs{Ultracold gases and trapped gases, Trapping of molecules, Molecular spectra}

\maketitle

The field of ultracold polar molecules has gained a forefront interest in recent years~\cite{Carr09}.~In contrast to the contact interaction between atoms, the electric dipolar interactions between polar molecules are long range and anisotropic. Together with their complex internal structure, polar molecules are promising candidates for investigating strongly correlated many-body systems~\cite{Baranov08,Pupillo08}, quantum information~\cite{DeMille02}, ultracold quantum chemistry~\cite{Krems08,Ospelkaus10}, and precision measurements~\cite{Zelevinsky08}. The recent breakthrough in the production of a near quantum degenerate gas of KRb molecules~\cite{Ni08} has paved the way for studying ultracold polar molecules with high phase-space density. Chemical reactions controlled by quantum statistics of the fermionic KRb molecules have recently been observed~\cite{Ospelkaus10}. Dipolar interaction effects on these molecular collisions have also been studied~\cite{Ni10}.

Convenient and powerful detection methods are at the heart of successful research with ultracold atoms~\cite{Ketterle99}. Optical cycling transitions make it possible for each atom to scatter many photons, which greatly enhances the signal-to-noise ratio (SNR) for optical absorption imaging. Images of the cloud can be taken with or without TOF expansion and these can be used to obtain the spatial and momentum distributions of the trapped gas. In the case of molecules, the wavefunction overlap between ground and excited states (Franck-Condon factor (FCF)) is typically much less than unity. The absence of cycling transitions then directly implies that the maximum number of photons scattered from the molecular target basically equals to the number of molecules. Hence, detection of a small number of cold molecules formed, for example, via photo-association typically relies on very sensitive ionization detection methods~\cite{Wang04a}. Laser-induced fluorescence detection on decelerated molecules has also been successfully employed~\cite{Bochinski03,Meerakker05}. Although these techniques are powerful, it is challenging to make their complex setups compatible with the ultrahigh vacuum systems required for quantum gas experiments. In addition, these methods need pulsed lasers to excite or ionize the molecules, and velocity imaging techniques would need to be implemented~\cite{Heck95} to obtain momentum information. In our previous work~\cite{Ospelkaus10, Ni10}, ground-state KRb molecules were detected using a complex approach of coherent transfer of the molecules to a weakly bound state near a Fano-Feshbach resonance followed by atomic absorption imaging. Because of the tiny binding energy, these molecules could be photodissociated and detected using light on the atomic cycling transitions~\cite{Ospelkaus06,Zirbel08}.

In this Letter, we demonstrate imaging of a high phase-space-density sample of ground-state polar molecules using direct optical absorption. The imaging is accomplished via an open optical transition~\cite{Gibble92,Mhaskara07}. The optical density is a dynamic quantity here since it decreases rapidly during the probe time as the population decays to many dark states. In addition, the absorption cross section for the molecule is usually at least one order of magnitude smaller than those for atoms, due to the low FCF and rotational branching. Here we implement absorption imaging on one of the strongest molecular bound-bound transitions for KRb. By counting the number of absorbed photons via a CCD camera, we obtain molecular imaging with a high SNR.  The highest SNR is achieved by using a minimum number of photons for imaging the molecules with photon shot-noise limited detection. The demonstrated technique can be used to determine momentum and spatial distributions for the molecular gas and allows for flexible molecular imaging at arbitrary external electric ($E$) and magnetic ($B$) fields.

\begin{figure}
\centering
\includegraphics[width=0.95\linewidth]{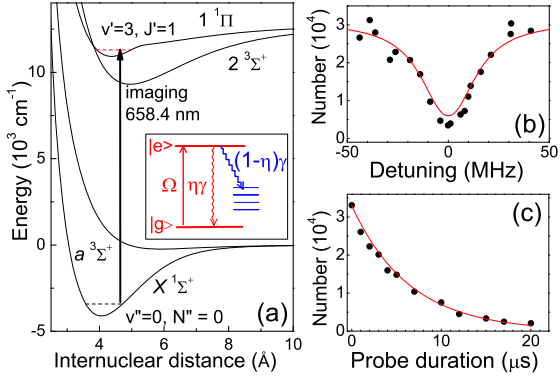}
 \caption{\protect\label{fig1} (Color online) (a) The $ab~ initio$ electronic potentials for KRb, showing the direct absorption detection scheme for $X^1\Sigma^+$ molecules in the $v''$ = 0, $N''$ = 0 level. The probe light drives a transition to the $v'$ = 3 level of the 1$^1\Pi$ state. Inset: Schematic diagram of the open two-level model described in the text. (b) An example population depletion lineshape for this transition taken with 10 $\mu$s probe pulse duration. (c) Decay of the ground-state population as a function of the probe pulse duration. The probe intensity is $\sim$0.7 mW/cm$^2$. The solid curves are fits based on the open two-level model.}
\end{figure}

Our experimental procedure to produce ultracold $^{40}$K$^{87}$Rb molecules in the lowest vibrational ($v''$ = 0) and rotational ($N''$ = 0) level of the $X^1\Sigma^+$ electronic ground state is described in Ref.~\cite{Ni08}. Near quantum-degenerate $^{40}$K and $^{87}$Rb atoms are prepared in a pancake-shaped crossed-beam optical dipole trap with an aspect ratio of 6 between the tighter confined vertical direction $z$ and the horizontal direction $x$. After creating weakly bound $^{40}$K$^{87}$Rb Feshbach molecules in a high vibrational state, we coherently transfer the molecular population to a single hyperfine state of the ro-vibronic ground-state manifold via stimulated Raman adiabatic passage (STIRAP)~\cite{Ni08,Ospelkaus10b}. Our standard method to detect these ground-state molecules has been to transfer them back into the weakly bound Feshbach state, followed by detection with the usual atomic absorption imaging method. Because of the coherent transfer process, this Feshbach imaging scheme can only detect molecules in a single hyperfine state. It is also very sensitive to external fields.

To image the ground-state molecules directly, we use a transition at $\lambda$ = 658.4 nm, from the absolute ground state to the vibrational $v'$ = 3, rotational $J'$ = 1 level of the 1$^1\Pi$ state, as shown in Fig.~\ref{fig1}~(a). The 1$^1\Pi$ state has been studied in detail for a different isotope combination, $^{39}$K$^{85}$Rb~\cite{Okada96}, and the selected transition is one of the strongest in the KRb molecule~\cite{Beuc06}. Using the available potential~\cite{Okada96}, we calculate the relevant rovibrational levels for $^{40}$K$^{87}$Rb. The measured transition frequency is within 30 GHz of the prediction. Shown in Fig.~\ref{fig1}~(b) is a typical molecular resonance lineshape taken at a relatively low intensity of $\sim$0.7 mW/cm$^2$. The x-axis is the probe laser frequency minus the measured transition frequency of 455219 GHz. The number of remaining molecules is measured using the Feshbach imaging method described above. The nature of an open transition can be seen in Fig.~\ref{fig1}~(c), where we show the loss of ground-state molecules as a function of the probe time. For these data, the probe laser frequency is on resonance.

\begin{figure}
\centering
\includegraphics[width=0.95\linewidth]{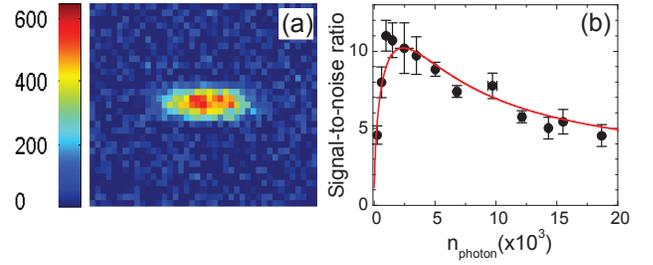}
 \caption{\protect\label{fig2} (Color online) (a) An absorption image (228 \textnormal{$\mu$}m $\times$ 171 \textnormal{$\mu$}m) of 39,000 ground-state molecules after 2 ms TOF. The false color indicates CCD counts, taken for the best SNR with $\sim$75\% of the molecules detected. (b) Peak SNR as a function of $n_{photon}$. Filled circles: experimentally measured peak SNR; Solid curve: fit to the model described in the text. The measured $N_{molecule}$ = 950 per pixel at the cloud center.}
\end{figure}

To calibrate the transition strength, we simultaneously fit several lineshapes and decay curves obtained for a range of  intensities to an open two-level model, which is shown schematically in the inset of Fig. 1~(a). This model is described by the optical Bloch equations (OBEs):	
\begin{eqnarray}\label{eq:image.1}
	\dot{\rho}_{gg} = \eta \gamma \rho_{ee}+\frac{i}{2} \Omega( \tilde{\rho}_{eg}- \tilde{\rho}_{ge})
	\nonumber \\
  \dot{\rho}_{ee} = -\gamma \rho_{ee}+\frac{i}{2} \Omega(\tilde{\rho}_{ge}- \tilde{\rho}_{eg})
  \nonumber \\
  \dot{\tilde{\rho}}_{ge} = -(\frac{\gamma}{2}+ i \Delta)  \tilde{\rho}_{ge}+\frac{i}{2} \Omega (\rho_{ee}-\rho_{gg}).
\end{eqnarray}
Here, the density matrix elements $\rho_{gg}$ and $\rho_{ee}$ are the fractional population in the ground and excited states, respectively. The other two elements $\tilde{\rho}_{eg}$ = $\tilde{\rho}_{ge}^*$ are related to the coherence. The laser frequency detuning from the transition is $\Delta$. The probe Rabi frequency is given by $\Omega =d \mathscr{E}/\hbar$, where $d$ is the electronic transition dipole moment including the FCF, $\mathscr{E}$ is the probe field and $\hbar$ is the Planck's constant over 2$\pi$. As the calculated FCF is 0.21~\cite{Okada96} and the rotational branching ratio is 2/3 according to the H{\"o}nl-London coefficients for the probe transition, the branching ratio $\eta$ is 0.14. Initially, all molecules are prepared in the ground state, $\rho_{gg}$ = 1 and $\rho_{ee}$ = 0. At $t > 0$, $\rho_{gg} + \rho_{ee} < 1$ due to the decay rate $(1-\eta)\gamma$ from the excited state to dark states. The excited population also decays back to the ground state with a smaller rate $\eta\gamma$. Shown by the solid curves in Fig.~\ref{fig1}(b) and (c) are the fits of the data to the OBEs with $\gamma$ and $\Omega$ as free parameters. From the fits, we determine $\gamma = 2\pi \times 22(7)$ MHz and $\Omega = 2\pi \times 0.8(1)$ MHz.

A typical direct absorption image of ground-state molecules is shown in Fig.~\ref{fig2}~(a). This image is obtained using a CCD camera. In one experimental cycle, two images are taken, one with molecules and the other with just the probe light for reference. By subtracting these two images, we obtain the molecular absorption image. Furthermore, the subtraction eliminates most of the technical background noise, which arises from spatial variations in the probe beam intensity. Taking into account the CCD calibration (including a quantum efficiency of 0.9 electron/photon and a digitizing gain of 1.1 electron/count), an 11$\%$ loss from the imaging optical elements, and the braching ratio $\eta$ = $0.14$, we need to multiply by a factor of 1.2 to convert CCD counts to molecule number. For each image, we perform a surface Gaussian fit to obtain the peak signal, $N_{peak}$, in CCD counts per pixel and sizes $\sigma_x$, $\sigma_z$ in number of pixels. The molecule number can then be obtained from $2 \pi N_{peak} \sigma_x \sigma_z \times 1.2$. The detected number of molecules by direct imaging is typically 15$\%$ lower than that measured by the Feshbach imaging method. To extract the ground-state molecule number from the Feshbach imaging method, we have to consider the STIRAP efficiency, the Feshbach molecule detection efficiency, and optical density saturation. Thus, the number estimate from direct imaging has smaller uncertainties. In addition, in contrast to conventional absorption imaging, the open-transition imaging eliminates the need for the exact value of the absorption cross section.

Since the total signal size is limited to the molecule number, it is crucial to understand the detection SNR and optimize it. We have verified that the SNR in our experiment is limited by shot noise, which scales with the square root of the photon number. Under the condition of weak probing ($\Omega << \gamma$), the fractional ground-state population follows $\rho_{gg}$ = $e^{-\Omega^{2} t (1-\eta)/\gamma}$. Here $t$ is the probe pulse duration and $\Omega^{2} t = 4 \pi \frac{d^2}{\hbar \epsilon_0 \lambda A}n_{photon}$, with $n_{photon}$ the average photon number per CCD pixel at the molecular cloud position. The area per pixel $A$ is 5.7 \textnormal{$\mu$}m  $\times$ 5.7~\textnormal{$\mu$}m at the cloud position and $\epsilon_0$ is the vacuum permittivity.  As each molecule can scatter $1/(1 - \eta)$ photons before decaying into a dark state, we express the signal size (total number of scattered photons) per pixel, $N_{signal}$, as a function of $n_{photon}$,
\begin{equation}\label{eq:image.2}
N_{signal} = \frac{N_{molecule}}{1 - \eta} (1-e^{-4 \pi \frac{d^2 (1-\eta)}{\hbar \epsilon_0 \lambda A \gamma}n_{photon}}),
\end{equation}
where $N_{molecule}$ is the initial number of molecules per pixel. The photon shot-noise-limited SNR is then
\begin{equation}\label{eq:image.3}
SNR = 0.8 N_{signal}/\sqrt{2 \times 0.8 \times n_{photon}},
\end{equation}
where the factors of 2 in the denominator arises from the fact that we subtract two frames to obtain one absorption image. The net detection efficiency is 0.8 (including losses from optics and quantum efficiency of the CCD).

Fig.~\ref{fig2}~(b) shows the peak SNR as a function of $n_{photon}$ where $n_{photon}$ is varied by using different $t$ at a constant probing intensity. The peak SNR is determined from the absorption image by comparing the measured number of counts per pixel at the cloud center to the standard deviation of the measured number of counts per pixel far from the cloud center. The solid line in Fig.~\ref{fig2}~(b) shows a fit to the model described by Eqns.~(\ref{eq:image.2}) and (\ref{eq:image.3}) with $d$ the only free parameter. The initial number of molecules per pixel at the cloud center, $N_{molecule}$ = 950, is determined from the data. The fit yields $d = 3.5(6)$ Debye, consistent with measurement of ref.~\cite{Okada96}. The SNR initially increases as $n_{photon}$ is increased; however, for large $n_{photon}$, the growth of photon shot noise dominates the signal increase. From Eqns.~(\ref{eq:image.2}) and (\ref{eq:image.3}), the best SNR occurs when 71$\%$ of the population is depleted by the probing light. In a typical experiment, we use $n_{photon}$ slightly greater than that for the optimal SNR to ensure that all molecules are depleted. This allows us to accurately count the total number of molecules.

\begin{figure}
\centering
\includegraphics[width=0.94\linewidth]{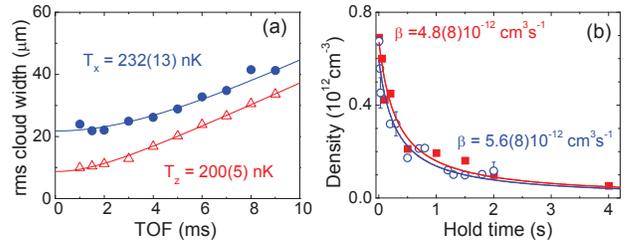}
 \caption{\protect\label{fig3} (Color online) (a) Measurement of the molecular gas temperature by TOF expansion followed by direct absorption imaging. The trap frequency along the $x$ and $z$ direction is 32 Hz and 195 Hz, respectively. Filled circles and open triangles are the rms cloud widths in the $x$ and $z$ directions, respectively. Solid curves are fits to a ballistic expansion. (b) Measurement of the molecular two-body loss rate $\beta$, for an initial temperature of 320 nK, using the direct imaging (open circles) and Feshbach imaging (filled squares) methods. Solid curves are fits to a two-body loss model. The extracted values of $\beta$ agree within the experimental uncertainty.}
\end{figure}

Fig.~\ref{fig3}~(a) demonstrates the use of direct absorption imaging to measure the temperature of a molecular cloud using TOF expansion following sudden release from the optical trap. We measure molecular cloud sizes for expansion times up to 9 ms. As the molecular cloud expands, the detection SNR decreases. For longer expansion times, we average several successive images to improve the SNR. From the measurements in Fig.~\ref{fig3}~(a), we extract a mean cloud temperature of 220(13) nK, which corresponds to $\sim$1.4 times the gas Fermi temperature.

We have also taken a lifetime measurement of the trapped molecules to study inelastic collisions. In Ref.~\cite{Ospelkaus10}, the observed molecular two-body loss, detected by Feshbach imaging, was due to chemical reaction of KRb + KRb. The fact that the molecules were prepared in the lowest energy state, including the hyperfine structure~\cite{Ospelkaus10b}, led to the unambiguous observation of chemical reactions. If the KRb molecules are instead prepared in an excited hyperfine state, inelastic hyperfine-state-changing collisions are also a possible loss mechanism. With the Feshbach imaging method, we cannot distinguish between the two possible loss channels because the imaging is only sensitive to a single hyperfine state. In the direct imaging method, the probe transition linewidth is much larger than the entire span of $\sim$ 3 MHz of the ground-state hyperfine manifold at 550 Gauss. The direct imaging thus provides a simultaneous detection of all the molecules regardless of their distribution among different hyperfine states. Hence, a comparison of direct imaging and Feshbach imaging could allow us to check if nuclear spin flips to other hyperfine states occur in an ultracold molecule gas. Fig.~\ref{fig3}~(b) presents exactly such a comparison. The data taken here are for ground-state molecules prepared in an excited hyperfine state. At a temperature of 320 nK, the loss rate coefficients obtained using the two imaging techniques are in good agreement. While this agreement is consistent with there being no appreciable rate for hyperfine-state changing collisions, at the large $B$ field used here, it is also possible that spin-flipped molecules might acquire a sufficient kinetic energy to leave the trap before imaging.

\begin{figure}
\centering
\includegraphics[width=0.97\linewidth]{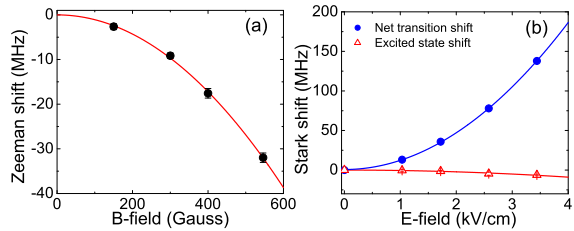}
 \caption{\protect\label{fig4} (Color online) Measurement of the Zeeman (a, black filled circles) and Stark (b, blue filled circles) shifts of the optical transition used for direct absorption imaging. Open triangles correspond to excited state Stark shifts obtained after subtracting the known Stark shift of the ground state~\cite{Ni08}; Solid curves are quadratic fits.}
\end{figure}

In future work, it will be interesting to explore molecule loss rates for different $B$ or $E$ fields. Here, the direct molecular imaging method is advantageous because it can operate under any combinations of $E$ and $B$ fields. Besides investigating the possibility of nuclear spin flips, the technique could also be useful in looking for molecular scattering resonances induced by $E$~\cite{Avdeenkov03} or $B$ fields. In addition, an important feature of direct imaging is the capability of imaging the molecular gas~\textit{in situ} under the presence of strong dipolar interactions.

In Fig.~\ref{fig4}, we show that the probing transition for direct molecule imaging is relatively insensitive to $B$ and $E$ fields. The $v'$ = 3, $J'$ = 1 level of the $^1\Pi$ state has first-order Zeeman and Stark shifts due to a half Bohr magneton magnetic dipole moment and a non-zero electric dipole moment. To minimize variation in the resonant frequency for probing over a broad range of $E$ and $B$ fields, we use the $M_J'$ = 0 sublevel as the excited state for detection. In doing so we eliminate the first-order Zeeman and Stark shifts in the excited state as they are both proportional to $M_J'$. Figure~\ref{fig4} shows the observation of relatively small (a) Zeeman and (b) Stark shifts over a large range of $B$ and $E$ fields. The quadratic dependences of the small shifts are indicative of the second order Zeeman and Stark effects, respectively. The $X^1\Sigma^+$ ground state has a negligible magnetic dipole moment, due only to the nuclear spin. Thus the observed Zeeman shift in Fig.~\ref{fig4}~(a) arises solely from the excited state. The Stark shift of the ground state was measured previously~\cite{Ni08}. The excited-state Stark shift can be determined by subtracting the ground-state Stark effect from the total observed shift (Fig.~\ref{fig4}~(b)).

In conclusion, we have demonstrated direct absorption imaging of an ultracold molecular gas. Whereas the previous technique of coherent transfer followed by absorption imaging of Feshbach molecules is highly state selective, direct imaging allows us to probe molecules independent of their hyperfine state. The detection scheme also eliminates the extreme sensitivity on $B$ and $E$ fields, enabling investigation of dipolar effects under various combinations of external fields. The detection scheme should be universal for all alkali dimers due to their spectroscopic similarities. We note that a non-destructive phase-contrast imaging technique~\cite{Andrews96} might be particularly advantageous for open structured molecules.

Funding is provided by NIST, NSF, AFOSR-MURI, and DARPA. *Present address: Norman Bridge Laboratory of Physics 12-31, Caltech, Pasadena, California.
**Present address: Max Planck Institute for Quantum Optics, Garching, Germany.

%\bibliography{molecule}% Produces the bibliography via BibTeX.

\end{document}